\documentclass{aa}  

\usepackage{graphicx}
\usepackage{txfonts}
\usepackage[breaklinks=true]{hyperref}
\hypersetup{
     colorlinks   = true,
     citecolor    = blue
}
\usepackage{makecell}

\begin{document}

   \title{High-resolution, high-sensitivity, ground-based solar spectropolarimetry with a new fast imaging polarimeter}

   \subtitle{Part I: Prototype characterization}

   \author{F. A. Iglesias \inst{1}, A. Feller\inst{1} \and K. Nagaraju \inst{1,2} \and S. K. Solanki \inst{1,3} }
   
   \institute{Max Planck Institute for Solar System Research,
   	   Justus-von-Liebig-Weg 3, 37077 G\"ottingen, Germany\\             
              \email{iglesias@mps.mpg.de}
         \and
            Indian Institute of Astrophysics, Koramangala Second Block, Bengaluru-560034, India
	\and
	School of Space Research, Kyung Hee University, Yongin, Gyeonggi-Do, 446-701, Korea.
             }

   \date{Received February 2, 2016; accepted March 30, 2016}

 
  \abstract
  {Remote sensing of weak and small-scale solar magnetic fields is of utmost
    relevance for a number of important open questions in solar physics. This requires the acquisition of spectropolarimetric data with
    high spatial resolution ($\sim10^{-1}$ arcsec) and low noise
  ($10^{-3}$ to $10^{-5}$ of the continuum intensity). The main limitations to obtain these measurements from the ground, are the degradation of the image
    resolution produced by atmospheric seeing and the seeing-induced crosstalk (SIC).}
  {We introduce the prototype of the Fast Solar Polarimeter (FSP), a new
    ground-based, high-cadence polarimeter that tackles the above-mentioned
    limitations by producing data that are optimally suited for the
    application of post-facto image restoration, and by operating at
    a modulation frequency of 100 Hz to reduce SIC. 
}
  {We describe the instrument in depth, including the
    fast pnCCD camera employed, the achromatic modulator package, the main calibration steps,
    the effects of the modulation frequency on the levels of
    seeing-induced spurious signals, and the effect of the camera properties on the image restoration quality.}
  {The pnCCD camera reaches 400 fps while keeping a high duty cycle
    (98.6 $\%$) and very low noise (4.94 $e^-rms$). The modulator is optimized to have high ($>80\%$) total polarimetric efficiency in the visible
    spectral range. This allows FSP to acquire 100 photon-noise-limited, 
    full-Stokes measurements per second. We found that the seeing induced signals present
    in narrow-band, non-modulated, quiet-sun measurements are (a) lower than
    the noise ($7\times10^{-5}$) after integrating 7.66 min, (b)
    lower than the noise ($2.3\times10^{-4}$) after integrating 1.16 min and (c) slightly above the noise ($4\times10^{-3}$) after restoring
    case (b) by means of a multi-object multi-frame blind
    deconvolution. In addition, we demonstrate that by using only narrow-band images (with low SNR of 13.9) of an active region, we can obtain one complete set of high-quality restored measurements about every 2 s.}
   {}
   \keywords{instrumentation: polarimeters -- techniques: polarimetric --
     techniques: image processing -- Sun: magnetic fields}

\titlerunning{Characterization of a novel high-cadence solar polarimeter}
\authorrunning{F. A. Iglesias et. al.}
\maketitle

%
\section{Introduction}
\label{sec:int}

Answering  many of the currently open questions in solar physics requires full-Stokes, imaging
spectropolarimetric measurements with high spatial (sub-arcesec), spectral
($<100$ $m\AA$) and temporal ($< 10 s$) resolution along with low
noise ($10^{-3}$ to $10^{-5}$ of the continuum intensity). Such science questions concern
e.g. the total amount of magnetic flux on the Sun \citep[e.g.][]{trujillo2004}, the existence and
significance of a potential small-scale turbulent dynamo \citep[e.g.][]{vogler2007,buehler2013,lites2014}, the structuring and
dynamics of the chromospheric magnetic field \citep[e.g.][]{solanki2003,harvey2009}, the contribution of weak
internetwork fields to solar irradiance \citep[e.g. ][]{solanki2013}, or magnetic processes related to
energy transfer through the solar atmosphere. The interaction of the magnetic
field and turbulent convection at the base of the solar atmosphere, structures
the field on scales spanning many orders of magnitude in size. There is no
doubt today that the plasma processes and magnetic field topologies on scales
below 100 km determine the energetics of the higher layers in the solar
atmosphere, through the chromosphere and transition region, up to the
corona. To acquire a better quantitative understanding of these processes, the
small-scale magnetic field has to be accurately measured, which leads to the
challenging requirements mentioned above.

Independently of an observatory's location, on the ground or in space, there are intrinsic trade offs between these requirements that arise from the fact that these kinds of measurements are photon starved. The latter being true even for telescopes, post-focus instrumentation and science cameras with the best possible photon-efficiency (i.e. highest throughput, quantum-efficiency and duty cycle).

The main hurdle to obtaining diffraction limited spatial resolution measurements in visible wavelengths from the ground, are the optical aberrations introduced by the always present Earth atmosphere (seeing). It is mainly the developments during the last 20 years in terms of adaptive optics (AO) systems \citep[reviewed in detail by][]{rimmele2011} and post-facto image restoration, that have allowed ground-based solar imaging to overcome seeing limitations and reach, although not with the same reliability, a quality comparable to space and balloon based observations \citep{kosugi2007,barthol2011}. The two aforementioned techniques, being complementary to each other, have become a standard in most of the state-of-the-art solar facilities and have partially motivated the construction of the next generation of large-aperture ($\ge$2 m) solar telescopes \citep[see e.g. review by][]{kleint2015}. 

The most common image restoration methods applied to ground-based solar data, are Speckle Imaging \citep{keller1992} and different variants of Multi-Object, Multi-Frame, Blind Deconvolution (MOMFBD) see \cite{lofdahl2002} and references therein. Independently of the technique used, the typical decorrelation time-scale of daytime seeing, sets the maximum exposure time of the individual images to some 10 ms \citep{lofdahl2007}. The low signal-to-noise ratio (SNR) of such short-exposure frames, makes the restoration of narrow-band images more difficult. This situation can be improved by using an extra high-SNR wideband (WB) channel synchronized to the narrowband science channel \citep[e.g.][]{vannoort2006,keller1992}, or by significantly increasing frame rate and duty cycle of the science channel, as will be shown in Sect. \ref{sec:ir}.  
    
In terms of polarimetric sensitivity, the main cause for the instrument to depart from its ideal photon-noise limited behavior is the detector noise. The latter can be reduced by minimizing dark current, i.e. cooling the sensor, and by adopting low-readout-noise electronics (see Sect. \ref{sec:hcpol}).  In addition, different spurious polarization signals can harm the measurement accuracy. These include the, sometimes combined \citep[e.g.][]{salmeida1994}, effects introduced by instrumental polarization,  by detector-related artifacts, and by atmospheric seeing or instrument jitter \citep{lites1987,vonderluhe1988}.

Instrumental polarization can be partially reduced either by modeling \citep[e.g.][]{beck2005b} or compensation \citep{stenflo1994,ramelli2014}. However only the latter helps to fight detector-related artifacts that scale with the level of instrumental polarization, e.g. sensor non-linearities or residual uncorrected offsets \citep{keller1996}. Moreover, some specific detector artifacts can also be reduced by accurate calibration and modeling (see Sect. \ref{sec:dacq}). On the other hand, there are mainly three methods used in ground-based polarimetry to cope with seeing induced crosstalk (SIC), which can severely harm the accuracy of the inferred magnetic field \citep[e.g. ][]{leka2001}. These are briefly outlined below.

(1) AO systems are known to reduce SIC \citep{judge2004, casini2012}. However, the residual artifacts can be large due to uncorrected high-order aberrations, finite bandwidth of the control system and seeing anisoplanatism. For example, when assuming a single-beam polarimeter modulating at 30 Hz, a Fried parameter of 10 cm and an ideal AO system compensating the first 30 Zernike terms \cite[see e.g.][]{roddier2004}, the numerical simulations of \cite{krishnappa2012} give for the main crosstalk component, from Stokes I to Q, U and V, a value of about $10^{-3}$ after 1 s integration time. For the crosstalk components between Q, U and V the authors found similar values.

(2) The combination of (slow) temporal and spatial modulation, particularly in the form of dual beam systems \citep{lites1987,collados1999}, which has successfully been applied to eliminate \textit{only} the main SIC component \cite[e.g.][]{collados2007, scharmer2008,socnav2006}, and gain a factor of $\sqrt{2}$ in SNR. This at the expense of introducing further spurious signals due to the differences in the beam path of each channel, e.g. flat field differences, differential image aberrations,  etc. Such beam imbalance artifacts are generally of order $10^{-3}$, with some stable systems reaching few times $10^{-4}$, e.g. \cite{beck2009} and \cite{lites2008}. One option, not very common in solar polarimetry, to further eliminate beam imbalance in dual-beam systems is the usage of the beam exchange technique \citep{semel1993,bianda1998,bommier2002}. This technique, however, when used on slow dual-beam systems, has its limitations in terms of high-resolution polarimetry, as it involves combining images recorded at different points in time, which might in turn result in a significant loss of spatial resolution due to seeing.

(3) Modulating much faster than the seeing to practically ''freeze'' the atmosphere within a modulation cycle period \citep{lites1987}. With most of the seeing power contained in the 1-100 Hz \citep{judge2004} frequency range, a modulation frequency of order 100 Hz is required to drastically suppress SIC \citep{krishnappa2012}.

An efficient modulation regime up to 1 kHz is reachable with electro-optical modulators based on ferro electric liquid crystals \citep[FLCs,][]{gandorfer1999}. FLCs are commonly used for full-Stokes polarimetry in the visible or near-infrared. These modulators have a proven longterm reliability and high optical quality, provided that thermal and UV radiation effects are taking into account. Moreover, total polarimetric efficiencies \citep[see][for a definition]{deltoroiniesta2000} above 80$\%$ are commonly obtained, usually by means of a modulator design optimized for efficiency across a broad spectral working range \citep{gisler2005,tomczyk2010}. 

Piezo-elastic modulators \citep[PEMs,][]{stenflo1985} operate at modulation frequencies of order 10 kHz where SIC becomes completely negligible. PEMs can also be used in the UV below about 380 nm, which is difficult to reach with liquid-crystal based modulators. However, a true full-Stokes PEM-based polarimeter with the required frequency stability has proven difficult to achieve in practice, see e.g. \cite{gandorfer1999}. Further, the extremely high modulation frequency, which is dictated by the acoustic resonance frequencies of the PEMs, rules out any demodulation by means of a synchronized frame readout, and requires other,  much faster demodulation techniques (see below).

The detection of kHz-modulated intensity signals has been the bottleneck of fast imaging polarimeters since the late 1980's, basically due to the inability to get large, 2D detectors with the required cadence and low noise \footnote{An alternative approach, based on optical demodulation of the intensity signals before detection,  eliminates the necessity of fast sensors \citep{stenflo1985}. However, this technique has not been applied because it involves losing a factor of three in intensity and other practical drawbacks like the alignment of three detectors, phase-locked modulator and demodulator, and many optical surfaces.}.  An exception being ZIMPOL \citep[e.g.][]{povel1990,gandorfer2004}, the only imaging polarimeter to this day being able to work in the kHz to 10 kHz range. ZIMPOL uses a dedicated CCD detector which allows on-chip accumulation of the different modulation states and therefore decouples the signal detection from the sensor readout \citep{povel1995}. This solution renders ZIMPOL fast enough to work with PEMs and has successfully been applied to obtain very high sensitivity ($<10^{-5}$)  measurements. However ZIMPOL is not the optimal instrument when high spatial resolution is also a goal (see Sect. \ref{sec:hcpol}).

In this work, we describe the prototype of a new, FLC-based, high-cadence, imaging polarimeter developed at the Max Planck Institute for Solar System Research in close collaboration with the semiconductor laboratory of the Max Planck Society. The Fast Solar Polarimeter (FSP) operates in the visible part of the spectrum and was conceived to reach the aforementioned challenging parameter domain, required for high-resolution, high-sensitivity solar polarimetry from the ground. This is possible thanks to a novel high-cadence camera that allows FSP to exploit the benefits of short exposure time, low detector noise, high duty cycle and fast modulation frequency with synchronous frame readout. 

With FSP we will be able to seriously tackle the science questions mentioned
above. For example FSP will be able to measure the magnetic flux in the quiet
Sun via deep Zeeman vector measurements in photospheric lines at about 2-3
times higher spatial resolution than in the infrared lines observed with other
polarimeters such as the Tenerife Infrared Polarimeter (TIP). Due to its
high sensitivity, FSP can also be used to detect spatial variations of the
Hanle depolarisation in the Sr I 460.7 nm line. This will test the hypothesis
that the Hanle depolarization of the Sr I line is mainly produced in the
intergranular lanes \citep{trujillo2004}. By carrying out the first spatially
resolved observations of this effect it might be possible to settle the
validity of the large amounts of magnetic flux (up to 160 G on average in the
quiet Sun, \citealt{shchukina2011}) previously found using this line.

To address the issue of a small-scale turbulent dynamo, FSP can follow up on
the technique pioneered by \citet{danilovic2010}, i.e. compare the spatial
distribution of Zeeman polarisation in high resolution observations with that
in small-scale turbulent dynamo simulations. Sensitive FSP Zeeman data in the
photospheric Fe I line pairs at e.g. 525 nm and 630 nm will allow extending
this technique to also comparing the evolution with time, which will further
strongly constrain the origin of the magnetic features.

Energy transfer through the solar atmosphere is another broad topic where we
expect significant contributions from FSP. FSP will be able to look for rapid
changes in photospheric and chromospheric magnetic fields, so far undetected
due to lack of appropriate instruments. This is crucial to detect high
frequency waves as well as the rapid evolution of magnetic fields in
flares. Furthermore, FSP can follow all magnetic features within its field of view
to determine how effectively field-line braiding takes place. This requires
Stokes vector observations at high resolution and low noise to catch also the
weak-field features. Otherwise the amount of braiding will be underestimated,
since features will be missed. FSP is predestined to provide the required
data.

The description of FSP is divided in two practically independent reports. This work ---hereafter refereed to as Paper I--- presents the instrumental details and \cite{feller2015} ---hereafter referred to as Paper II--- communicates the results of the first measurements obtained with the FSP prototype.

The present paper is structured as follows. Sect. \ref{sec:hcpol} compares the main properties of the FSP detector with respect to the ones used in other visible solar polarimeters to emphasize the need for a high-cadence, low noise sensor. Sect. \ref{sec:char} describes the instrument and presents the main performance characterization results including those of the modulation package (Sect. \ref{sec:modchar}) and the detector (Sect. \ref{sec:camchar}). Sect. \ref{sec:dacq} details the data acquisition and the main calibration procedures. Sect. \ref{sec:sic} briefly discusses the measured behavior of SIC at different modulation frequencies. Finally Sect. \ref{sec:ir} presents an example of the application of MOMFBD to FSP data.
 
\section{Detectors used in state-of-the-art, solar polarimeters}
\label{sec:hcpol}
 
 As explained in Sect. \ref{sec:int}, the intensity detector is a critical component to get the desired properties of a high-spatial-resolution, high-sensitivity,  ground-based polarimeter.  We have compiled in Table \ref{tab:scomp} nine important, detector-related properties of the following state-of-the-art, full-stokes, ground-based polarimeters that work in the visible part of the spectrum\footnote{We excluded from this list the SOLIS/VSM \citep{keller2003} although it is an interesting example of early high-cadence polarimetry. The initial idea of the project in 1998 was to use a pair of custom made, 1024x1024, Sarnoff VCCD1024H cameras which had, 18x18 $\mu m^2$ pixel area, 46 $e^-rms$ readout noise and 300 fps frame rate \citep{keller1998}. However the system was never delivered by the manufacturer, forcing the VSM team to select an alternative camera, with the Rockwell HyViSI-1024 model being chosen. The cameras were modified to have 256x1024 pixels, a 92 fps frame rate and have been operative since 2003 \citep{harvey2004}. Due to variable dark-levels, inter-quadrant crosstalk and readout noise issues, among others, VSM was updated in 2010 to use a pair of commercial Sarnoff 1M100 cameras \citep{balasu2011} similar to the ones used in SPINOR, hence the exclusion from our list.}:  IBIS \citep{cavallini2006}; VIP \citep{beck2010}; CRISP \citep{scharmer2008, rodriguez2015}; DLSP on phase II \citep{sanka2004, sanka2006}; SPINOR \citep{socnav2006} ;  FSP (see Sect. \ref{sec:char} and \ref{sec:dacq}) and ZIMPOL-3  \citep{ramelli2010,ramelli2014}. We included, in addition, the expected properties of the FSP project in its second phase, labeled FSPII (see Sect. \ref{sec:cncl}).
  
  \begin{table*}[!htbp]
 \centering
\caption{Approximate values of the main, detector-related properties for eight state-of-the-art, full-stokes, ground-based, solar polarimeters working in the visible part of the spectrum. Instruments marked with ``*'' operate in spectrograph mode while the others are used in filtergraph mode. FSP has been tested in both modes. See the text for details.}
\label{tab:scomp}
  \begin{tabular}{l l c c c c c c c c c }		
  \cline{1-11}   
   \noalign{\smallskip}  
 & Polarimeter  & Pixel area & Detector & Frame rate & Ro. noise  & ADC & Duty & Mod. &   Dual  & WB \\
 	&&  $[\mu m^2]$ &  area $[px^2]$ & $[fps]$&$[e^-rms]$&  $[bits]$ & cycle $[\%]$ &  freq. $[Hz]$ & beam & channel  \\
  \noalign{\smallskip}
  \cline{1-11}   
   \noalign{\smallskip}     
 & IBIS  & 6.8$\times$6.8 &1024$\times$1024 & 2.86 & 20 & 12 & 3\tablefootmark{a}   & 0.42 & Yes & Yes \\   
 & VIP      & 16$\times$16 & 512$\times$512 & 29& 5.4 & 16 & 100 & 7.25 & Yes & Yes \\  
 & CRISP  & 16$\times$16 & 1024$\times$1024 & 37 & 20 & 12 & 70 & 9.25 & Yes  & Yes \\ 
 & DLSP*  & 12$\times$12 & 488$\times$652 & 50& 50 & 14 & 100 & 12.5 & Yes & No  \\   
 &SPINOR*& 16$\times$16 &  512$\times$1024 & 100 & 40 & 12 & 100 & 12.5 & Yes & No \\  
 &  FSP & 48$\times$48 & 264$\times$264 & 400\tablefootmark{b}  & 4.9 & 14 & 98.6 & 100 & No& No  \\
 &  FSPII\tablefootmark{c}  & 36$\times$36 & 1024$\times$1024 & 400& 5 & 16& 95 & 100 & Yes & Optional\tablefootmark{d}  \\
 &  ZIMPOL-3*\tablefootmark{,e} & 22.5$\times$90 & 1252$\times$144& 1.47& 6 & 16 & 1.5\tablefootmark{a}   & $\ge1000$ & No& No\\  
 \noalign{\smallskip}       
 \cline{1-11}
\end{tabular}
\tablefoot{
   \tablefoottext{a}{Computed for an exposure time of 10 ms.}
  \tablefoottext{b}{The FSP camera can achieve a frame rate of 1100 fps using a special readout mode. However, we consider the maximum to be 400 fps because it is the value used during the observations presented here and in Paper II. Such value is also the expected maximum frame rate for FSPII.}
  \tablefoottext{c}{Currently under development, we listed the expected values. See Sect. \ref{sec:cncl}.}
  \tablefoottext{d}{The camera noise is less critical in the WB channel due to the higher flux, thus, it can be implemented with e.g. a fast commercial CMOS camera.}
    \tablefoottext{e}{Even though the CCD55-30 sensor used has frame transfer architecture, the ability of reading while integrating is not implemented resulting in an overhead readout time of about 0.66 s.}
 }
\end{table*}
Note that, for some of the  polarimeters listed in Table \ref{tab:scomp}, the adopted values of the detector properties may differ from the maximum allowed by the cameras. The latter is the result of design trade-offs that may include the performance of associated post-focus instruments or different detector parameters, e.g. operating the camera at a reduced frame rate to minimize readout noise. In all cases we listed the values that are reported to be used during typical polarimetric observations. We excluded from Table \ref{tab:scomp} the optical filling factors, with most of the sensors having almost 100$\%$; and quantum efficiency, because in general it can be modified by means of sensor coatings in order to guarantee high (>80$\%$) values in the desired portions of the visible spectrum.
 
 In the following list, we briefly discuss all the properties presented in Table \ref{tab:scomp}, along with their effect on imaging polarimetric performance, and compare their values for the different instruments:
  \begin{enumerate}
 \item \textit{Pixel area}: Two aspects are relevant regarding pixel area. Firstly, detectors with non-square pixels have a more complex point spread function (PSF) and sample the image at different spatial frequencies in the two orthogonal dimensions, further complicating the image restoration process. Secondly, the pixel size has to match the required spatial sampling of the ---ideally diffraction limited---  image. As a consequence, large pixels may require re-imaging setups of unpractical dimensions, which can also introduce further optical aberrations.
 
Therefore, small and square pixels are an advantage. From the second column of Table \ref{tab:scomp}, note that this aspect constitutes a drawback of the ZIMPOL-3 concept, namely the usage of a sensor with three out of four rows covered,  implies the attachment of a micro-lenses array in front of the photosensitive matrix for focusing the light on the uncovered pixels and keep near 100$\%$ filling factor. The latter leads to non-square pixels with one large dimension of 90 $\mu m$ --- almost a factor of six larger that the most common values (<16 $\mu m$) in Table \ref{tab:scomp}--- and the necessity of a challenging sub-micron, alignment process between pixels and micro lenses \citep{gandorfer2004}. Besides ZIMPOL-3, the next largest pixel areas, two to three times larger than the other listed polarimeters, belong to FSP and FSPII, this is mainly due to limitations of the manufacturing process and may limit their usage in some solar facilities.
 
 \item\textit{Detector area}: The size and aspect ratio of the detector, directly affect the instrument FOV. In general, large (1k $\times$ 1k) detectors are required to image most common scientific targets with sub-arcsec sampling, e.g. active regions. In combination with a spectrograph, a large aspect ratio can be an advantage, depending on the required spectral resolution and range. For filtergraph-based systems, smaller aspect ratios are commonly used.
 
There are a variety of sizes and aspect ratios listed in the third column of Table \ref{tab:scomp}, with FSP having the smallest dimensions. Large detector dimensions, high frame rate and low readout noise constitutes a general trade-off that depends among others on the readout strategy and technology. Cameras that use massive parallel readout are more easily scalable to larger sensor areas, while keeping similar cadence and readout noise, than those based on serial strategies. This is the case for the pnCCD (See Sect. \ref{sec:camchar}) sensors used in FSP and FSPII, with the small 264$\times$264 and the, nearly four times larger, 1024$\times$1024 versions being able to reach the same high frame rate (400 fps), low readout noise ($\leq$5 $e^-rms$) and very similar duty cycles ($\ge$95 $\%$).  

 \item \textit{Frame rate}: A high frame rate is needed to get a short exposure time ---required to apply image restoration--- without sacrificing duty cycle, and a high modulation frequency to reduce SIC levels (see Sect. \ref{sec:int}). 

As can  be appreciated in the fourth column of Table \ref{tab:scomp}, FSP can reach four times (400 fps) the frame rate of the next fastest polarimeter, SPINOR (100 fps). The latter implies, given the modulation scheme adopted (see Sect. \ref{sec:modchar}), that FSP can take a full Stokes measurement in 10 ms, making it optimal for image restoration and for high-cadence polarimetry of strong signals (see Sect. \ref{sec:ir} and Paper II).
 
 \item \textit{Readout noise}: Readout noise is usually the dominant noise source in cooled detectors. If photon-noise limited observations are a goal, when working with low photon fluxes, then adopting a low readout noise camera is of paramount importance. The latter is what discourages the usage of commercial high-speed (>kHz) cameras to do high-resolution polarimetry.
 
From the figures given in the fifth column of Table \ref{tab:scomp}, it can be seen that VIP, ZIMPOL-3, FSP and FSPII have readout noises approximately one order of magnitude smaller that the rest of the listed instruments. However, among these, FSP and FSPII have more than 13 times higher frame rate.  
 
 \item \textit{ADC}: The longer the analog to digital converter (ADC) word, the lower the quantization noise ---typically included in the readout noise figure of the camera---  added during the digitalization of the sensor voltages \citep{bennett1948}. Note that, a real ADC will introduce an extra amount of noise and distortion, due to its non ideal circuitry, that increases with the conversion rate. The latter is specified by the ADC effective number of bits or ENOB \citep{IEEE2001}, which is smaller than the ADC word length (reported in the sixth column of Table \ref{tab:scomp}). For high-speed, high-dynamic-range cameras, the trade-off  between digitalization noise and sample rate of the ADC may become an issue, e.g. for a sensor with 50,000 $e^-$ full well capacity and a 12 ENOB ADC, the digitalization noise is 3.5 $e^-rms$ \citep[][Sect. 4.2.5]{hoslst1998}. 

  \item \textit{Duty cycle}:  High-resolution, spectropolarimetric observations of dynamic signals in the Sun are photon starved, therefore utilizing every photon that reaches the detector is crucial. The duty cycle, i.e. the product of the frame rate times the exposure time, is then a determinant parameter. Furthermore, the main motivation to use detectors with full frame transfer architecture (see Sect. \ref{sec:camchar}) in polarimetry, is their almost 100$\%$  duty cycle even at short exposure times. 
  
The difference between detectors with and without frame transfer architecture is clearly seen in the seventh column of Table \ref{tab:scomp}. Sensors that are not full-frame transfer, as for IBIS, or that do not make use of the reading-while-integrating capability of this architecture, as for ZIMPOL-3, have overhead readout times that make them almost unusable for polarimetry at both, high spatial resolution and high sensitivity due to the very low resulting duty cycle. 
 
 \item \textit{Modulation frequency}:  As mentioned in Sect. \ref{sec:int}, the detection of the modulated intensities can be done in synchronization with the detector readout, or not. The latter case, requires a special sensor and is the corner-stone of the ZIMPOL systems. For all the other polarimeters given in Table \ref{tab:scomp}, the modulation frequency is mainly limited by the detector frame rate. Different modulation schemes and technologies require different number of intensity measurements, N, to obtain the full Stokes vector. The modulation frequency is given by the detector frame rate divided by N; consequently, modulation schemes with low N are to be preferred in this respect \footnote{For example, a fast-rotating waveplate polarimeter \cite[e.g.][]{hanaoka2012}  has N=8, and thus requires twice the detector frame rate of an FLC-based system, which has N=4, to achieve the same modulation frequency.}. All the synchronous-readout polarimeters listed have N=4 except for IBIS that has N=6 and SPINOR that has N=8. 
 
 Column number eight of Table \ref{tab:scomp} details the varied modulation frequency values of the selected polarimeters; with ZIMPOL-3 almost four order of magnitudes  faster than IBIS and ten times faster than the FSPs respectively. Note that the high-frame-rate of the FSP and FSPII cameras and the N=4 modulator, allow these polarimeters to modulate eight times faster than the next fastest polarimeter that uses synchronous readout (SPINOR), further reducing the SIC levels to very low values (see Sect. \ref{sec:sic}).
 
\item \textit{Dual beam}: As can be appreciated in column number nine of Table \ref{tab:scomp}, the dual beam technique is commonly used to reduce SIC in slow polarimeters, being not necessary for kHz modulation frequencies. The performance of a fast (>50 Hz) dual beam system, strongly depends on the relative values of the residual beam imbalance artifacts and SIC. This point has not yet been addressed in other solar polarimeters and thus motivates the implementation of an optional dual beam setup to be used with FSPII, with the two orthogonal polarization images illuminating different sections of the sensor, i.e. halving the FOV (see Sect. \ref{sec:cncl}).
 
\item \textit{WB channel}: As for the case of a dual beam system, acquiring WB images simultaneously to the polarimetric narrow-band data, requires either reducing FOV or the installation of an extra camera with the same frame rate. This is also the case for simultaneous, multi-wavelength observations. In the aforementioned situations, the monetary costs may become an issue if custom-made, expensive detectors are used in the polarimeter. 
\end{enumerate}

In the present section we have compared only cameras that have been tested in solar polarimeters, which are mostly CCD based. However, very-recently-introduced, scientific CMOS cameras (e.g. from manufacturers such as Andor, Fairchild and Hamamatsu), yield very promising specifications in terms of readout noise, frame rate and quantum efficiency, for their application in solar polarimetry (provided that the non-linearities, fixed pattern noise and any artifact relevant to differential photometry can be controlled).
\section{Instrument description and characterization}
\label{sec:char}

\subsection{System overview}
\label{sec:so}
\begin{figure*}[!htbp]
\centering
\includegraphics[scale=0.9]{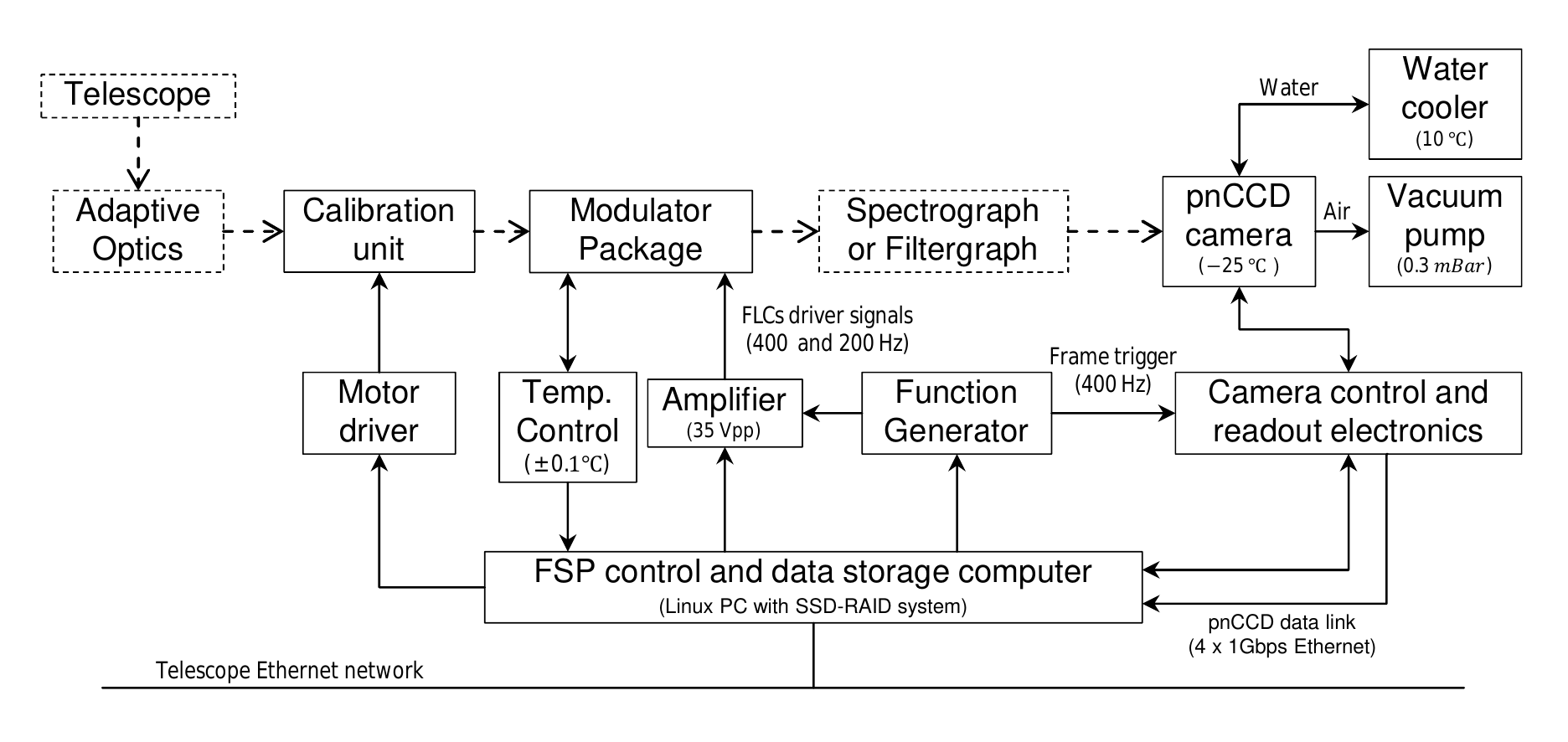}
\caption{Components of the FSP prototype. Dashed arrows indicate the beam path. Non-labeled connections using solid arrows represent different control, data and power lines. Typical operating parameters for some of the blocks are annotated in parentheses. The dashed blocks do not belong to FSP. Two extra communication lines, not shown here, are required to synchronize the spatial or spectral scanning with FSP acquisition, when a filtergraph or spectrograph is used, respectively.}
\label{fig:block}%
\end{figure*}

The different components of the FSP prototype are represented in the block diagram of Fig.\ref{fig:block}. The various blocks and connections are detailed in the following sections. As can be appreciated, the polarimeter operates in a single-beam configuration. The light (dashed arrows) coming from the output port of the AO system passes through the FLC-based, modulator package ---first passing through the instrument calibration unit  (ICU) if necessary--- to enter the wavelength discriminator system, which can be either a spectrograph or a filtergraph. After this, there is a synchronous detection of the modulated intensities using the pnCCD camera. The signals used to trigger frame acquisition and to drive the modulator are generated by a laboratory function generator.  Since the FLCs use a voltage of 35 Vpp for a correct switching, a two-channel laboratory amplifier is used. The complete system is commanded by a single Linux-based computer that, in addition, records the data during measurements (See Sect. \ref{sec:dacq}).
    
\subsection{Polarization modulation}
\label{sec:modchar}

\begin{figure}[!htbp]
\centering
\includegraphics{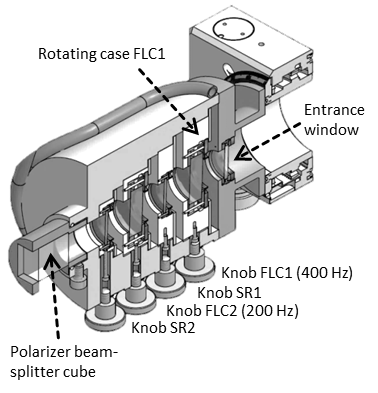}
\caption{CAD cross-section of FSP modulator package. The main components are two Ferro-electric Liquid Crystal (FLC), half-wave retarders (FLC1 and FLC2); a quarter-wave, static retarder (SR1); a half-wave, static retarder (SR2); and a polarizer beam splitter cube. The two FLCs and SRs are mounted inside rotating cases to permit the manual modification ---by means of the four labeled knobs at the bottom--- of their optical axes orientation. The latter allows the optimization of the efficiencies for a given wavelength and/or Stokes parameter in case a specific scientific program requires it.}
\label{fig:mod}%
\end{figure}

The FSP modulator package is similar to the SOLIS/VSM design \citep{keller2003}. The ideal monochromatic version consists of two FLC, half-wave retarders (FLC1 and FCL2); a quarter-wave, static retarder (SR1); a half-wave, static retarder (SR2);  and a polarizer beam-splitter cube, used as linear analyzer. The two bistable FLCs combined, provide four different modulation states and, thus, the same number of intensity measurements are necessary to obtain the full Stokes vector, i.e. N=4. A CAD cross-section of the modulator is presented in Fig. \ref{fig:mod}. 

The temperature dependence of the FLCs switching angles and retardances \citep{gisler2003}, demands a thermal stabilization of the modulator assembly to within a fraction of a degree. The large-mass housing and active thermal control, provided by a closed-loop system (see Fig. \ref{fig:block}) based on a flexible pad heater, ensures a temperature stability of $\pm$0.1 $^\circ C$. As a result, a highly stable modulator was obtained, e.g. polarimetric efficiencies that were measured three days apart from each other did not differ by more than 1.5 $\%$. The latter significantly reduces the overhead for polarimetric calibration.

The modulator was conceived to minimize the chromatism and unbalance, i.e. the difference between Stokes Q, U and V, of the polarimetric efficiencies in the 400-700 nm wavelength range. Since the manufacturing tolerances of the FLCs can be large, e.g. $\pm10\%$ for the retardances, we applied the following design procedure in order to minimize their effect:
 \begin{itemize}
\item In a first step, two half-wave FLCs were purchased from Boulder Nonlinear Systems and subject to careful characterization. At a given temperature, the dispersion of the retardance, $\phi(\lambda)$, can be expressed in fractions of the wavelength, $\lambda$, as \citep{gisler2003}:
\begin{equation}
\phi(\lambda) = \left(\frac{\lambda_0}{2}-\frac{C_d}{\lambda_0^2}\right)\frac{1}{\lambda}+\frac{C_d}{\lambda^3},
\label{eq:disp}
\end{equation}

where, $\lambda_0$ and $C_d$ are the central wavelength and the dispersion constant respectively. 

\item The measured switching angles, $\lambda_0$ and $C_d$ for the FLCs (see Table \ref{tab:modp}), were used as input to obtain a first optimal set of position angles for all the components, and of retardances for the SRs. The optimization procedure, developed by \cite{gisler2005}, simultaneously minimizes, within the 400-700 nm range, the squared differences between the polarimetric efficiencies of a modulator model and the ideal ones for a perfectly balanced and achromatic system, i.e. constant and equal to 1, 0.58, 0.58, 0.58 for Stokes I, Q, U and V respectively \citep{deltoroiniesta2000}.

\item After this, the SRs were ordered from the manufacturer, specifying the desired retardances obtained in the previous step. 
\item Finally, $\lambda_0$ and $C_d$ of the actual SRs purchased were measured  (see Table \ref{tab:modp}) and used to run a second optimization with the position angles as the only free parameters to fit.
 \end{itemize}
\begin{table}[!htbp]
 \centering
\caption{Properties of the optical components in FSP modulator. $C_d$ and $\lambda_0$ are model parameters fitted to reproduce the measured dispersion of the retardances at 40 $^\circ C$  (see text). The position angles are the optimum to maximize achromatism and balance of the polarimetric efficiencies in the 400-700 nm wavelength range.}
\label{tab:modp}
  \begin{tabular}{l l c c c c c}		
  \cline{0-6}   
   \noalign{\smallskip}  
 &Property &FLC1 &SR1 &FLC2 &SR2 & Unit \\  
  \noalign{\smallskip}
  \cline{0-6}   
   \noalign{\smallskip} 
 &Retardance\tablefootmark{a}&0.482&0.283&0.453&0.576&$\lambda$  \\ 
 &Switching ang.\tablefootmark{a}&41.5&-&42.6&-&$\circ$  \\ 
 & $C_d$ $(x10^7)$&1.41&0.25&1.41&0.46&$nm^3$ \\ 
 & $\lambda_0$&451.9&510.8&434.2&517.7&$nm$  \\ 
 &Position ang.\tablefootmark{b}&$ -71.8$&$ 26.7$&$ -41.5$&$ 64.8$&$\circ$  \\ 
  \noalign{\smallskip}     
 \cline{0-6}
 \end{tabular}
\tablefoot{
   \tablefoottext{a}{Measured at 460 nm and 40 $^\circ C$.} 
   \tablefoottext{b}{Defined with respect to the optical axis of the polarizer beam-splitter cube, for the transmitted beam.}
  }
\end{table}

\begin{figure*}[htb]
\centering
\includegraphics[scale=0.634]{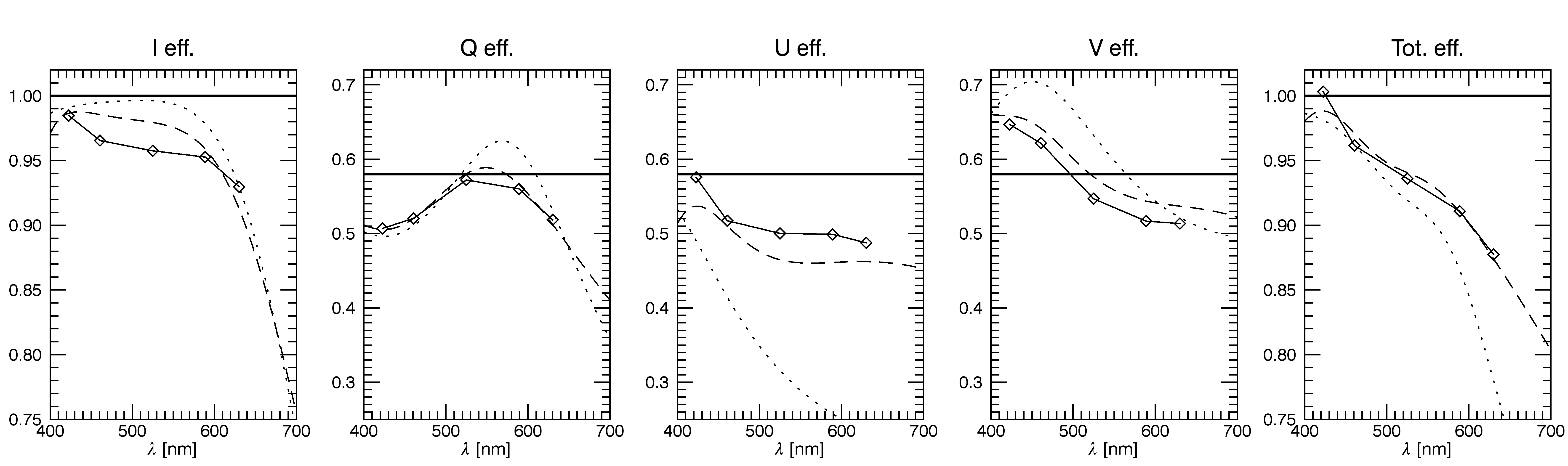}
\caption{Measured (diamonds), ideal (thick continuous) and modeled FSP polarimetric efficiencies vs. wavelength. The results of the numerical modulator model are presented for two different configurations, one using the values listed in Table \ref{tab:modp} (dashed) and the second using the same input except that the values of the dispersion coefficients, $C_d$, were all set to zero (dotted). The large differences of the latter with respect to the measured values, particularly in Stokes U, emphasize the necessity of a detailed model of the retardances' dispersion in order to properly reproduce the spectral response of the modulator. The ideal efficiencies are the maximum achievable by a perfectly balanced and achromatic system.}
\label{fig:eff}%
\end{figure*}

 The final properties of the optical components in the FSP modulator and the optimal position angles found,  are summarized in Table \ref{tab:modp}. The measured spectral response of the polarimetric efficiencies is shown in Fig. \ref{fig:eff} along with the results of the numerical modulator model used to perform the optimizations, obtained when the values in Table \ref{tab:modp} are input. We also include in Fig. \ref{fig:eff} the results of the modulator model after setting all the $C_d$ constants to zero (dotted line). This is done to emphasize the necessity of a detailed modeling of the retarders' dispersions, i.e. Eq. \ref{eq:disp}, in order to perform a correct optimization procedure.

\subsection{pnCCD camera}
\label{sec:camchar}

The pnCCD camera was custom-made by PNSensor GmbH together with the MPG semiconductor lab, and is the main component of FSP. It provides simultaneously high frame rate, high duty cycle and low readout noise (see Table \ref{tab:scomp}), allowing the polarimeter to reach a modulation frequency of 100 Hz with synchronous readout. This is accomplished by using a back-illuminated, 3-phase, pnCCD sensor that has a split frame-transfer architecture. Furthermore, all the sensor columns are readout in parallel by means of the 528  amplifiers ---integrated in four readout chips--- that use multi-correlated double sampling to keep the readout noise low. The measured camera non-linearities are approximately 1$\%$ and can be reduced to below 0.1$\%$ after calibration. Therefore, polarimetric effects due to non-linearities \citep{keller1996} are in the order of $10^{-5}$ given an instrumental polarization of order $1\%$. The quantum efficiency of the sensor is above 90 $\%$ in the 500-850 nm wavelength range, and above 60 $\%$ within 390-1000 nm. Further details on the camera structure and performance are given in \cite{ihle2012} and \cite{hartmann2006}.

The measured readout plus quantization noise at 400 fps is 4.94 $e^-rms$ with a standard deviation over the sensor area of $0.16$ $e^-rms$, for a conversion gain of 8.68 digital counts per electron. At the normal camera operating temperature of -25 $^\circ C$, the measured dark noise is $0.31$ $e^-rms$. The camera is cooled down using a water-refrigerated (see Fig. \ref{fig:block}), thermo-electric Peltier device. As a consequence, part of the camera housing is evacuated (see Fig. \ref{fig:block}) down to 0.3 mBar to avoid water vapor condensation on the sensor and electrical contacts.

\section{Data acquisition and calibration}
\label{sec:dacq}

 FSP can be used in combination with a spectrograph or a filtergraph as post-focus instrument for wavelength discrimination (see Fig. \ref{fig:block}). In both cases, the final outcome is a data cube composed of two spatial and one spectral dimensions for each of the four Stokes parameters. Filtergraphs ---such as Fabry-P{\'e}rot interferometers, e.g. GREGOR/GFPI \citep{bellogonzalez2008} and VTT/TESOS \citep{kentischer1998,tritschler2002}---  are the option of choice in high-spatial resolution observations because the  spatial information is recorded strictly simultaneously and, within the isoplanatic patch dimensions of the seeing, degraded by the same PSF. As a consequence, filtergraphs can benefit from post-facto image restoration techniques. On the other hand, spectrographs ---typically slit-based--- can detect a full spectrum simultaneously and be used with long integration times to reach high polarimetric sensitivities, at the expense of reduced spatial resolution. A newly developed technique (M. Van Noort, private communication) promises to allow spectral scans to be restored as well. This would make the use of FSP with spectrographs of considerable interest in the near future.

Independently of the post-focus instrument, the data rate of FSP is large, namely 390 Mbits/s at 400 fps. Consequently four parallel Ethernet connections and a solid-state-disk (SSD) based RAID storage system are used to record the acquired data (see Fig. \ref{fig:block}).

\subsection{Detector calibration}
\label{sec:scal}
The following list describes, in order of application, the main data reduction steps used during the calibration of the measured intensities. The steps are independent of the post-focus instrument unless explicitly specified:
\begin{enumerate}

\item\textit{Offset}: The modulated intensity images recorded with the pnCCD camera are corrected for an offset by subtracting a low-noise dark frame, obtained by averaging a long dark series.

\item\textit{Common mode}:  A characteristic issue in parallel readout sensors, as the pnCCD, is the so called common mode artifact. This is a variable offset introduced during readout that has, in a given frame, approximately the same value for all the amplifiers belonging to a single readout chip. Therefore, for the pnCCD the artifact appears semi-row wise (see the left image in  Fig. \ref{fig:cm}). Common mode level dominates the image power in dark frames and changes randomly from exposure to exposure, thus it has to be corrected in order to perform accurate differential imaging. Such a correction is performed in FSP by subtracting, from each semi-row, the common mode signal estimated from the average value of seven out of eight shielded pixels located in the borders of the sensor ---the pixel closest to the illuminated area is not used due to the elevated stray light--- reducing the usable area from 264  to 248 columns. Fig. \ref{fig:cm} illustrates the results of the common mode correction in a dark frame. In addition, Table \ref{tab:cm} gives the values of the residual power in dark frames when different numbers of shielded pixels are used to estimate the artifact level.
\begin{figure}[h]
\centering
\includegraphics{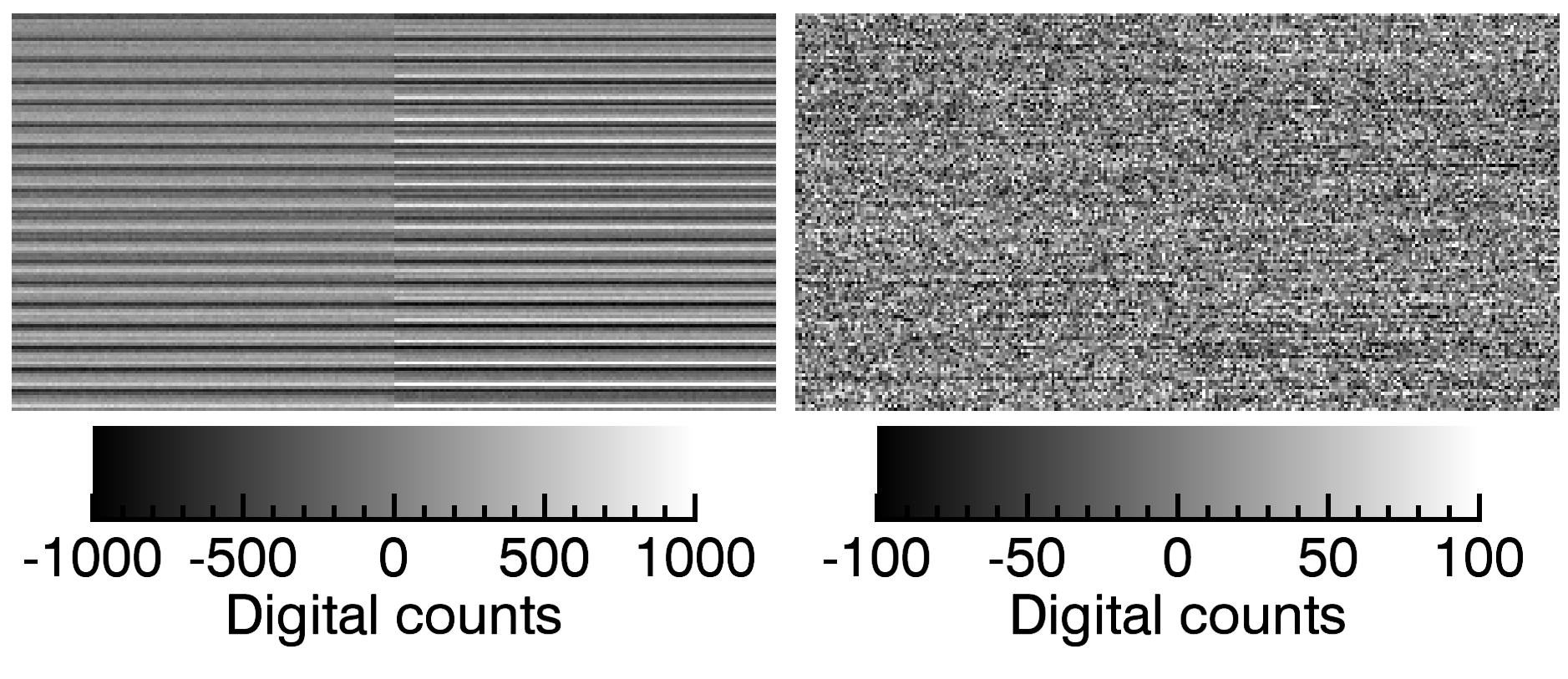}
\caption{Offset-corrected dark frame taken with the pnCCD camera, before (left) and after (rigth) the common mode correction. Only a fraction of the sensor area is shown. The common mode signals are estimated for each semi-row using seven shielded pixels, not shown here, located at the borders of the sensor. The images were taken at 400 fps and $-25\text{ }^\circ C$.}
\label{fig:cm}%
\end{figure}
\begin{table}[h]
 \centering
\caption{Measured total noise in dark frames for different numbers of shielded pixels used to correct the common mode artifact. Using zero pixels means that no correction is applied.  Both the mean values and the standard deviations of the noise, computed over the sensor area, are given in $e^-rms$. The frames were recorded at 400 fps and -25 $^\circ C$, with a conversion gain of 8.68 digital counts per electron.}
\label{tab:cm}
  \begin{tabular}{l c c c c c }		
  \cline{0-5}   
   \noalign{\smallskip}    
    Num. of pix. & 0 & 1& 3 & 5& 7  \\
    Total noise  & 45.32 &  6.54 &  5.35 &   5.07 &  4.94 \\
    Spatial $\sigma$  & 9.65 & 0.22 & 0.17 & 0.17 & 0.16 \\    
 \cline{0-5}
 \end{tabular}
\end{table}

\item\textit{Frame transfer}:  In order to keep the duty cycle as high as possible, FSP does not use a shutter. Since the sensor is permanently illuminated during the frame transfer, image smearing can appear \citep{hoslst1998}. Due to the high frame rate of the pnCCD camera, the frame transfer takes approximately 1.4$\%$ of the exposure time. Therefore, the resulting smearing artifact may be of concern for some high-contrast scenes, for this reason, a numerical technique was developed to perform a post-facto correction of the images for the general case of non-constant illumination of the sensor, see \cite{iglesias2015} for details.

\item\textit{Flat field}: Flat field correction is applied by dividing the four modulated intensities by the same offset, common mode and smearing corrected, normalized flat frame; i.e. polarized flat field effects are not taken into account. The flat frame is obtained by randomly moving the telescope around the solar disk center. When working with a spectrograph, the spectral lines are numerically removed from the flat frame post-acquisition.
\end{enumerate}

For filtergraph-based observations, after the above-described sensor calibration, the images can be optionally restored using image restoration techniques such as MOMFBD, this is detailed in Sect. \ref{sec:ir}. The measured intensities are then demodulated using the demodulation matrix obtained from the polarimetric calibration procedure explained in the following section.

\subsection{Polarimetric calibration}
\label{sec:cal}

To perform the polarimetric calibration of FSP, the motorized ICU is introduced in the beam path in front of the modulator package (see Fig. \ref{fig:block}). Note this calibration does not include instrumental polarization introduced by the telescope or the AO, these are addressed for a specific case in Paper II. The ICU is composed  of a high-quality linear polarizer and a super achromatic quarter-wave plate, i.e. retardance is between 77$^\circ$ and 93$^\circ$ in the 400-700 nm wavelength range.

To obtain the polarimeter demodulation matrix, 19 known polarization states are created with the ICU ---by rotating its motorized retarder  in 10$^\circ$ steps--- and measured by FSP. The unknown $4\times4$ elements of the (de-)modulation matrix are then fitted to the measurements. An example of a demodulation matrix obtained at a wavelength of 630.25 nm is presented in Table. \ref{tab:modm}. 

In order to avoid image shifts, introduced by the switching of the FLCs \cite[see e.g.][]{hanaoka2006}, the modulator package is located close to the scientific focal plane of the telescope. Thus, spatial variations of the demodulation matrix over the FOV are expected. Such variations introduce undesired fluctuations of the measured Stokes parameters across the image that scale with the value of the instrumental polarization. Consequently, two options to lessen this effect are (1) the reduction of the instrumental polarization \footnote{Which also has other advantages like the reduction of polarimetric artifacts introduced by sensor non linearities and residual offsets \citep{keller1996}.}, e.g. given the maximum (three sigma) variation in the FSP demodulation matrix is 2.7$\times10^{-2}$, the instrumental polarization level should be 1$\times10^{-2}$ in order to keep the polarization errors near 1$\times10^{-4}$ ; and (2) performing the polarimetric calibration and demodulation for each pixel independently.
\begin{table}[!htbp]
 \centering
\caption{Measured FSP demodulation matrix at a wavelength of 630.25 nm. For each element, both the mean and the standard deviation (in parenthesis) were computed over the sensor FOV and are expressed in $\%$.}
\label{tab:modm}
  \begin{tabular}{r r r r}		
  \cline{0-3}   
   \noalign{\smallskip}    
     34 (0.6)  &  36 (0.6)   &  19 (0.5)  &  11 (0.7)\\
    -23 (0.7)   & -54 (0.7)   &  1 (0.6)    &   76 (0.8)\\
     67 (0.8)   &  -45 (0.4)  &  36 (0.4)  &  -58 (0.9)\\
     33 (0.5)   &   27 (0.7) &  -82 (0.6) &   22 (0.6)\\    
 \cline{0-3}
 \end{tabular}
\end{table}

\subsection{Normalization and image accumulation}
\label{sec:iac}

After demodulation, the polarization levels are normalized with respect to the intensities, i.e. the images of Stokes Q, U and V are divided by Stokes I. Additionally, many individual Stokes images may be accumulated in order to increase SNR. The accumulation and normalization operations do not commute and thus the order of their application has to be selected.

 For example, in the case of Stokes Q, the two possible options can be expressed for each pixel in the images as $\langle Q/I\rangle $ and $ \langle Q \rangle / \langle I \rangle$, where $\langle\rangle$ denotes the average of $M$ images. Considering that both Q and I are noisy quantities with mean values equal to $\mu_Q$ and $\mu_I$,  respectively, one is interested on the best estimator of the ratio $\mu_Q/\mu_I$. Out of the two aforementioned estimators, only  $ \langle Q \rangle / \langle I \rangle$ is asymptotically unbiased for $M\to\infty$ \cite[e.g.][]{vankempen2000}. This can be appreciated after writing the second order approximation to their expected values as follows,
\begin{align}
 \mathbf{E}\left ( \left  \langle\frac{Q}{I}\right \rangle \right)=\frac{\mu_Q}{\mu_I}+bias && \text{and} &&   \mathbf{E}\left( \frac{\langle Q \rangle}{\langle I \rangle}\right)=\frac{\mu_Q}{\mu_I}+\frac{bias}{M},
\label{eq:expv}
\end{align}
where $bias=(\kappa \mu_Q-\sigma^2_{Q,I})/\mu_I^2$, $\mathbf{E}$ represent the expectations value operator, $\sigma^2_{Q,I}$ is the covariance between Q and I, and $\kappa=1$ Digital count for a Poissonian I. Since $\mu_I$ is in the denominator of $bias$, the resulting artifact is non-flat and increases with the frame rate.

\begin{figure}[!htb]
\centering
\includegraphics{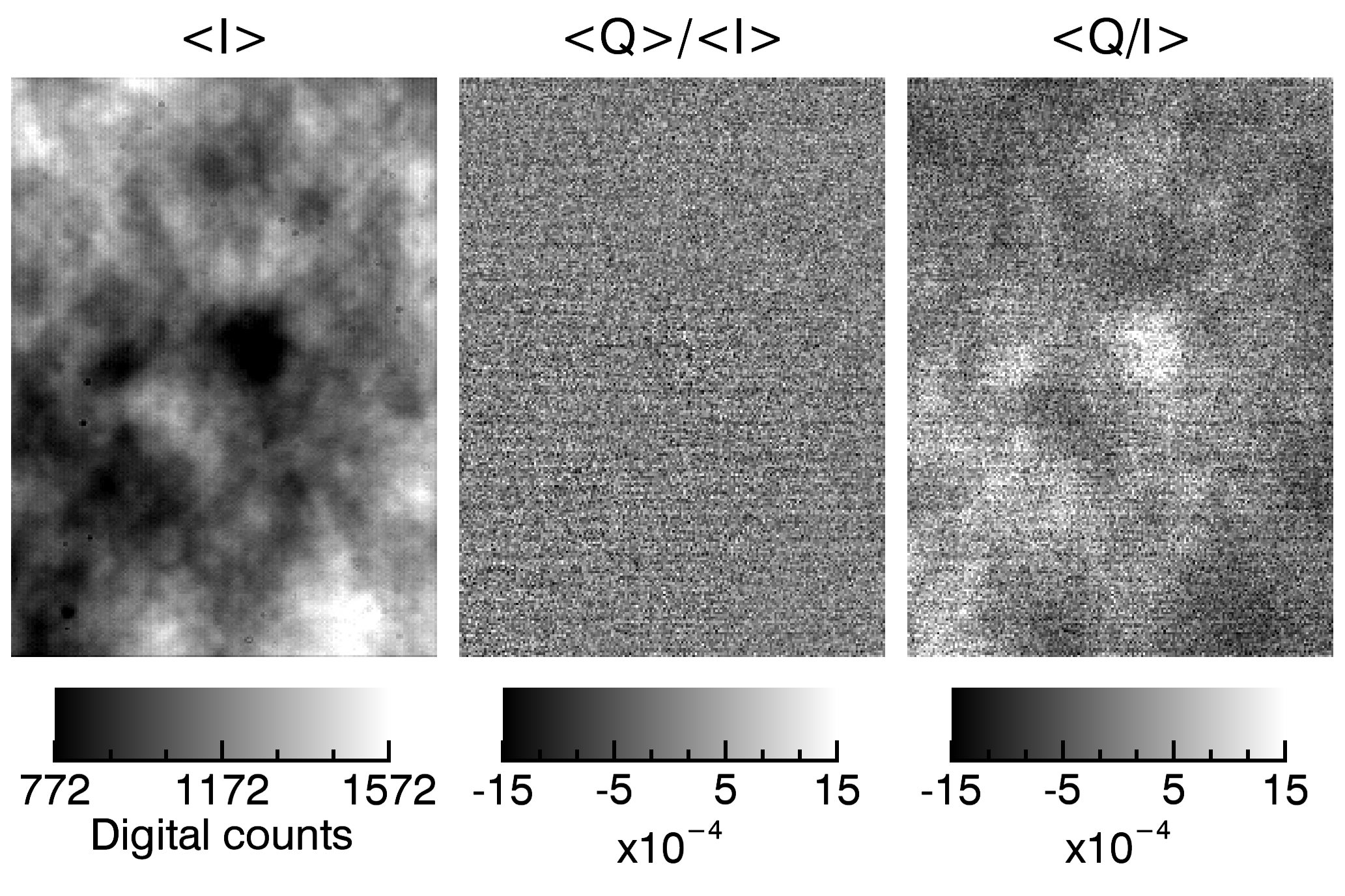}
\caption{Average of nearly 5$\times10^4$ (8.3 min) FSP measurements acquired with the modulator package disconnected. Only a fraction of the sensor area is shown. The contrast of the averaged, non-flat-fielded Stokes I image (<I>) is practically dominated by the etalons cavity errors. The difference of the artifact introduced by the $bias$ term of Eq. \ref{eq:expv}, is evident in the Stokes Q images that were accumulated and then normalized (<Q>/<I>) and vice versa (<Q/I>).}
\label{fig:norm}%
\end{figure}

The difference between the two estimators is exemplified by Fig. \ref{fig:norm} where approximately 5$\times10^4$ Stokes images ---corresponding to the measurements of a pore region acquired at -280 $m\AA$ from the line core of Fe I 630.25 nm--- were averaged. The images were recorded at 100 Hz modulation frequency using the VTT/TESOS filtergraph and with the FSP modulator deactivated. The latter implies that the Mueller matrix of the modulator is constant in time and thus no real polarization signal is expected\footnote{Even though no modulation is taking place, we still use the term modulation frequency to refer to one quarter of the detector frame rate. The same applies to Sect. \ref{sec:sic}}, i.e. $\mu_Q=0$ across the FOV.

\section{Spatial resolution and seeing-induced crosstalk}
\label{sec:sic}

Quantifying the amplitude of SIC in narrow-band, diffraction-limit-sampled, Stokes images recorded at high modulation frequency, is difficult because the artifact is normally hidden by the photon noise and any possible solar signal. Moreover if image averaging is used to reduce photon noise, the SIC artifact is smeared  and thus its level also reduces \citep[see e.g.][and Sect. \ref{sec:sicqs}]{hanaoka2004}. 

Let us discriminate the total power in the image of the normalized Stokes parameter $i\in\{Q,U,V\}$, after averaging M measurements as follows,

\begin{equation}
\sigma^2_{tot}(i,M,f,I_{rms})=\frac{\sigma^2_{Y}}{M\xi^2_i}+\sigma^2_{art}(i,M,f,I_{rms})+\sigma^2_{sig}(i,M),
\label{eq:totp}
\end{equation}
where we have neglected any fixed pattern noise that cannot be reduced by image averaging. $\sigma_{Y}$ is the noise in the measured intensities ---normalized to the mean intensity, divided by $\sqrt{N}$ and assumed to be the same in the four modulation states--- including the contributions from photon shot noise, dark shot noise and total readout noise; $\xi_i$ is the polarimetric efficiency of Stokes parameter $i$; $\sigma_{art}(i,M,f,I_{rms})$ includes the power due to SIC and other artifacts (see Sect. \ref{sec:sicqs}) in Stokes parameter $i$, where we made explicit its dependence with respect to $M$, to the modulation frequency, $f$, and to the rms contrast values of the Stokes I images averaged, $I_{rms}$; and $\sigma_{sig}(i,M)$ is the averaged signal power of the actual solar scene.

To be able to reliably detect solar polarimetric signals, one needs to obtain after averaging, a value of the signal to artifact and noise ratio, $SANR(i,M)=\sigma_{sig}(i,M)/[\sigma^2_{Y}/(M\xi^2_i)+\sigma^2_{art}(i,M,f,I_{rms})]^{1/2}$, that is large enough given a predefined criterion. Note that the above-described approach is of practical relevance only for the cases where the portion of the FOV covered by signal and by artifact are similar, so that their spatial rms values are equally representative of their peaks values.

In addition, to preserve spatial resolution, the desired $SANR(i,M)$ should be reached as fast as possible, i.e. for the smallest M, in order to avoid the reduction of $\sigma_{sig}(i,M)$ due to the spatial smearing produced by solar evolution and seeing. For a given photo-electron flux, the ideal case corresponds to a $\sigma_Y$ that is photon noise dominated and  $\sigma_{art}(i,M,f,I_{rms})/[\sigma_{Y}/(\xi_iM)]<<1$. As discussed in previous sections, the former is achieved for narrow band images only by adopting a very low noise camera. On the other hand, a low value for the SIC artifact requires a fast modulation frequency.

\subsection{Seeing induced crosstalk in quiet Sun images}
\label{sec:sicqs}

As an application of the above-expressed ideas, let us consider the case of quiet Sun measurements, of particular importance for FSP given its science goals. One way to achieve $\sigma_{sig}(i,M)=0$ in Eq. \ref{eq:totp} and still be sensitive to SIC, is to use measurements acquired with the modulator package deactivated (as in Sect. \ref{sec:iac}). In addition, since SIC not only depends on modulation frequency, but also on the  instantaneous seeing conditions and  contrast of the solar scene \citep{judge2004}; we use the following procedure to reduce the influence of the latter two.  

\begin{figure*}[!htb]
\centering
\includegraphics{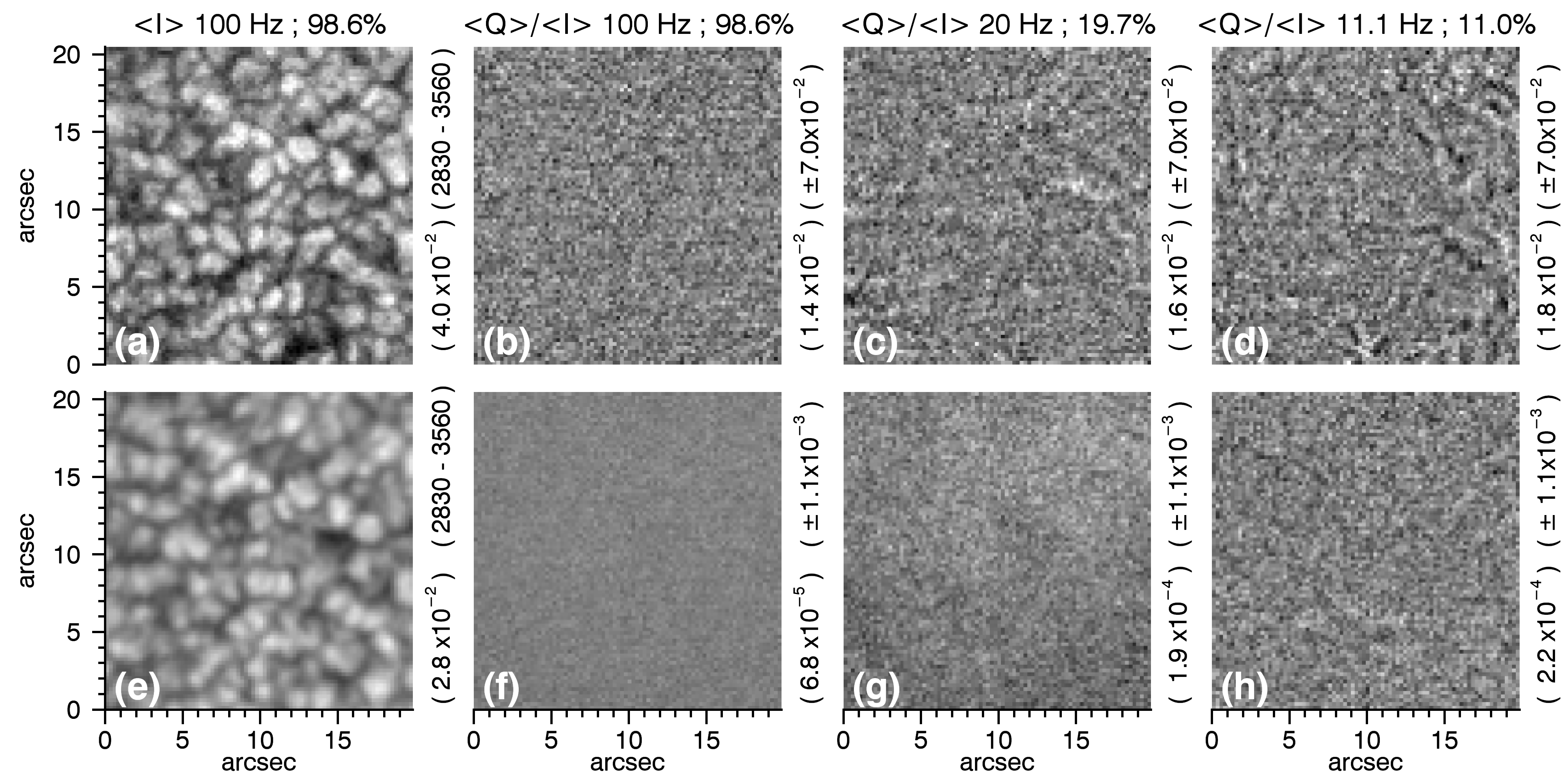}
\caption{Seeing induced crosstalk in a normalized Stokes Q image, for different modulation frequencies, duty cycles (labeled on top of each column) and number of averaged measurements (rows). These quiet Sun images were acquired with FSP modulator package disconnected, i.e. in which situation no solar polarization signal is expected, using the  VTT/TESOS filtergraph tuned to the continuum at -280 $m\AA$ from the line core of Fe I, 630.25 nm. The upper row presents Stokes I (a) and the normalized Stokes Q (b, c and d) of a single measurement. The lower row shows the results after averaging 45989 (e and f), 9160 (g) and 5060 (h) Stokes images, all covering 7.66 min wall time approximately. The 20 and 11.1 Hz cases,  were obtained by eliminating intermediate intensity measurements from the 100 Hz case before demodulation, see the text for details on this procedure. The original plate scale of the images is 0.08 arcsec/pixel in both directions, however they have been binned using a 3x3 $pixel^2$ window. The figures in parentheses on the right border of each image denote its gray scale range (upper figure) and standard deviation, or rms contrast for Stokes I, computed across the sensor area (lower figure).}
\label{fig:sic}%
\end{figure*}

We start with a set of 45989 (7.66 min wall time) quiet Sun measurements acquired at 100 Hz with the FSP modulator package deactivated. We then derived two new sets that have modulation frequencies of 20 and 11.1 Hz, by picking one out of five and one out of nine frames respectively, from the set of measured intensities at 100 Hz. After demodulation, the 20 and 11.1 Hz data sets are equivalent to measurements taken at those frequencies with duty cycles of 19.7$\%$ and 11.0$\%$ respectively. The advantages of this method are that practically the same solar scene, seeing conditions and values of $\sigma_{Y}/\xi_i$ are present in the three data sets. The measured values for Stokes I and normalized Stokes Q corresponding to the original and derived measurements are shown in Fig. \ref{fig:sic} for different values of M.  The qualitative results also apply to Stokes U and V, which are not presented here. 

Firstly, note that SIC is evident in the single measurements, i.e. M=1, of the 20 and 11.1 Hz cases (images c and d respectively) while not being above the noise level for the 100 Hz image (b). Moreover, the artifact tends to have larger amplitudes towards the  borders of the FOV because the AO corrections of the wavefront are worse away from the system's locking point ---located approximately at the center of the image--- due to seeing anisoplanatism \citep{rimmele2011}. 

\begin{figure}[!htb]
\centering
\includegraphics{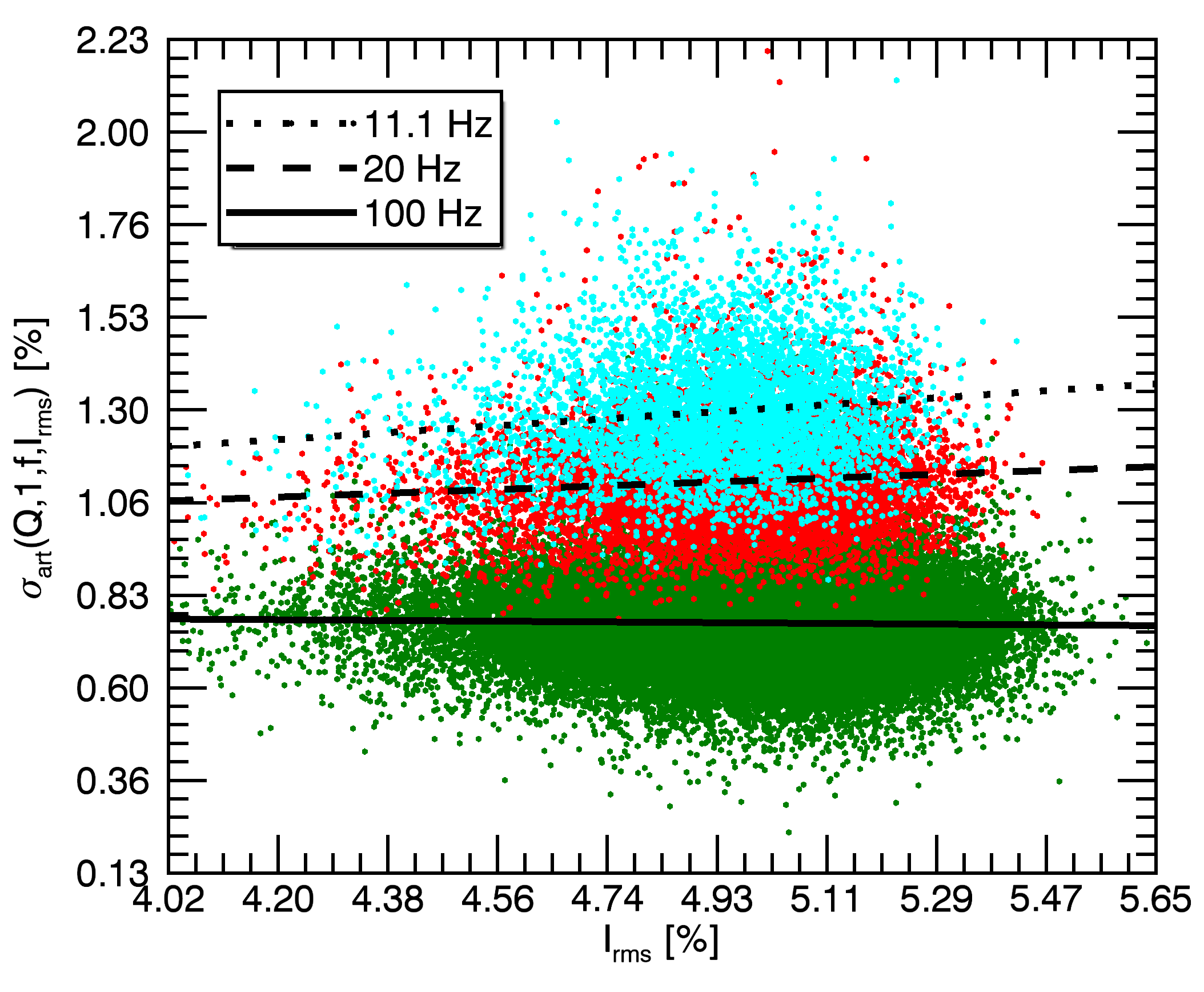}
\caption{Measured artifact level in individual Stokes Q measurements (see Eq. \ref{eq:alsq}) versus Stokes I rms contrast, for the data sets presented in Fig. \ref{fig:sic}. Each dot gives the value of a single measurement acquired at a modulation frequency of 100 (green), 20 (red) and 11.1 Hz (cyan). The black lines represent linear fits to the 100 (continuous), 20 (dashed) and 11.1 Hz (dotted) data sets.}
\label{fig:artl}%
\end{figure}

The reduction of the SIC with increasing modulation frequency, can be also appreciated in Fig. \ref{fig:artl}, where the artifact level in the individual Stokes Q measurements,
 \begin{equation}
 \sigma_{art}(Q,1,f,I_{rms})=\sqrt{\sigma^2_{tot}(Q,1,f,I_{rms})-\frac{\sigma^2_{Y}}{\xi^2_Q}},
 \label{eq:alsq}
 \end{equation}
is plotted for all the frames  ---thus including a variety of instantaneous seeing conditions--- in the three data sets versus Stokes I rms contrast. To obtain the artifact level, we assume it can only introduce power at spatial frequencies that are below the diffraction limit, and that $\sigma_{Y}$ is white noise. Then, $\sigma_{Y}/\xi_Q$ can be estimated by the average value of the power located at spatial frequencies beyond the diffraction limit. This estimation of $\sigma_{art}(Q,1,f,I_{rms})$ includes not only SIC contributions but also any other source of colored noise.

Secondly, from the second row in Fig. \ref{fig:sic}, note that SIC has been smeared out to very low levels in the 20 and 11.1 Hz cases (g and h respectively) due to image accumulation. This suggests that SIC artifact is less critical for low-spatial resolution measurements; where, in addition, any small-scale solar signal would also be smeared (compare the contrast of images a and e). If however, a high spatial resolution is aimed for, reducing the SIC below the photon noise level in the individual measurements is crucial\footnote{Note that this becomes more difficult when the photon flux reaching the detector increases, e.g. by taking sub-diffraction-limited images with a  larger aperture.}.

\begin{figure*}[!htb]
\centering
\includegraphics{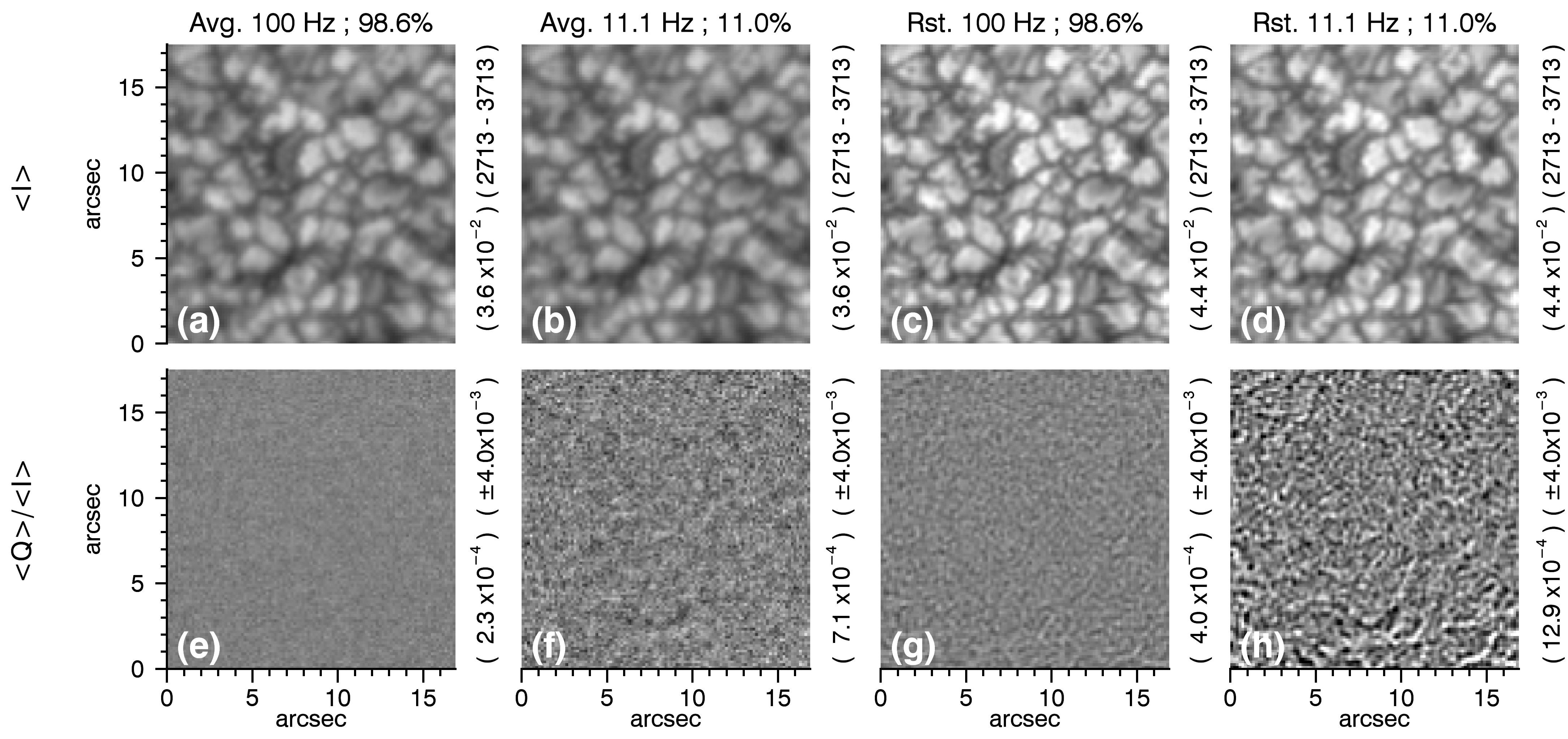}
\caption{Effects of the modulation frequency, duty cycle and image restoration (labeled on top of each column) on the SIC level in Stokes Q images. The Stokes I (upper row) and the normalized Stokes Q (bottom row) are shown for the data sets with 100 and 11.1 Hz  modulation frequency presented in Fig. \ref{fig:sic}. Both a plain average, covering 1.16 min wall time (a, b, e and f), and the corresponding  MOMFBD-restored versions (c, d, g and h) are presented. The original plate scale of the images is 0.08 arcsec/pixel in both directions, however they have been binned using a 2x2 $pixel^2$ window. The figures in parentheses on the right border of each image, denote its gray scale range (upper panels) and standard deviation, or rms contrast for Stokes I, computed across the images (lower panels).}
\label{fig:sic2}%
\end{figure*}

The latter is also exemplified by Fig. \ref{fig:sic2} were the resulting Stokes I and normalized Stokes Q, corresponding to the average of 1.16 min wall time of the data sets with 100 and 11.1 Hz modulation frequency, are presented along with their respective MOMFBD restored versions (see Sect. \ref{sec:ir} for details). In the plain averages, both the spatial resolution and the SIC levels are lower because signal  (images a and b) and artifact has been smeared out (images e and f). Thus, the benefits of a higher modulation frequency are less prominent. If image restoration is applied to preserve spatial resolution (see images c and d), and any possible small-scale polarimetric signal, then the SIC is also preserved producing a clear difference between the artifact levels of the cases with 100 and 11.1 Hz modulation frequency. The benefits of a combined high modulation frequency, high cadence and high duty cycle ---to obtain simultaneously high-spatial resolution (through image restoration) and low SIC--- are then demonstrated by images c and g.

\section{Image restoration}
\label{sec:ir}

In this section we present the application of the MOMFBD algorithm, developed by \cite{lofdahl2002} as an example of post-facto restoration of FSP data. In the multi-frame, blind deconvolution (MFBD) approach, both the object (the constant solar scene) and the degradation functions (the variable seeing plus instrumental PSFs) are simultaneously estimated assuming a noisy, linear shift-invariant degradation model in a maximum-likelihood fashion. In order to constrain the (initially ill-posed) problem and obtain a more unique solution, multiple short-exposure frames of the same object are recorded, the degradation functions are parametrized, and optical restrictions are added. 

The number of frames required for a MFBD is in principle small ($\sim$5), however it depends on the SNR of the individual frames and contrast of the solar scene \citep{lofdahl2007}. In typical narrow-band, short-exposure measurements, the photon noise is large compared to the signal and thus, if no WB channel is used, an increased number of frames is needed to get a satisfactory restoration. The latter becomes an issue when fast-evolving solar signals are considered, because the MFBD algorithm assumes a constant solar scene. In this respect, the duty cycle of the camera is critical in order to guarantee not only accumulating the maximum number of photons to get a higher SNR, but also a larger number of frames within the evolution time of the targeted solar feature at the desired spatial-resolution. 

Since FSP can record the full Stokes vector in 10 ms, i.e. below the typical daytime seeing decorrelation time scale ($\sim50$ ms), and with 98.6$\%$ duty cycle, the resulting polarimetric data can be reduced by (a) restoring the modulated intensities and then performing the demodulation, or by (b) restoring directly the normalized Stokes images. Both approaches have advantages and disadvantages and are suitable for different measurement regimes. 

The following is an example of approach (a) used to obtain a high-cadence (covering 1.92 s) restored image from 192 Stokes measurements acquired with FSP. The restoration was performed using the MOMFBD code implemented by \cite{vannoort2005}. The four groups of modulated intensities, each containing 192 frames, was input as a separate, incomplete object for the algorithm to simultaneously fit independent degradation functions for each modulation state. 

\begin{figure*}[!htb]
\centering
\includegraphics{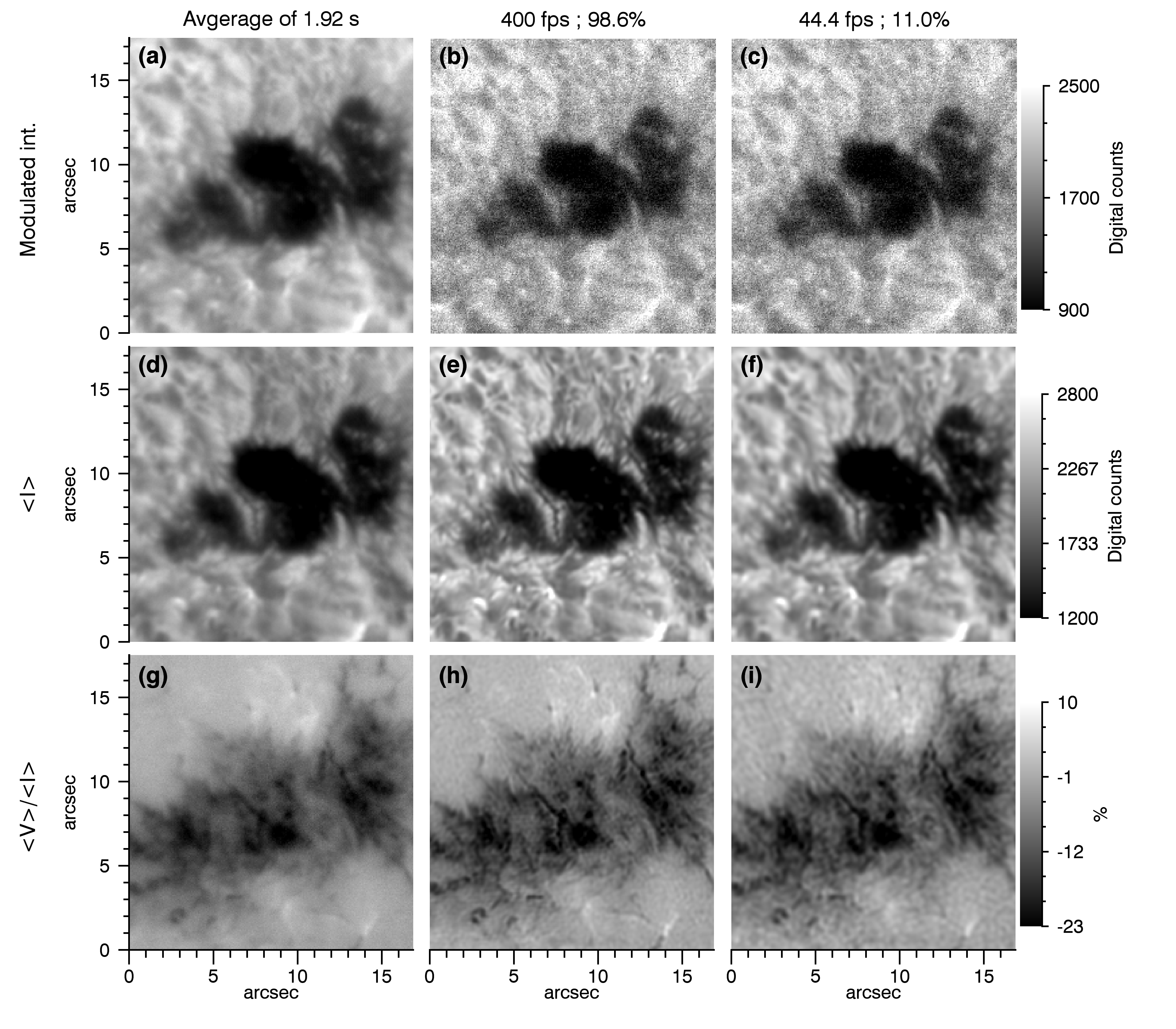}
\caption{Effects of the polarimeter cadence and duty cycle when restoring 1.92 seconds of narrow band data, by means of a MOMFBD. The first column presents a simple average of the 192 measurement cycles including, one modulated intensity (a), Stokes I (d), and normalized Stokes V (g). The second column illustrates the results of the MOMFBD restoration performed using all the $4\times192$ available intensity measurements. Both a single modulated intensity (b) and two of the resulting Stokes parameters (e and h) are shown. The third column gives the outcome of the same MOMFBD algorithm, run on the data set that has a reduced frame rate (44.4 fps) and duty cycle (11.1$\%$), including a single modulated intensity (c) and two of the restored Stokes parameters (f and i). The data with 44.4 fps frame rate was obtained by eliminating intermediate intensity measurements from the original FSP data set (acquired at 400 fps and a duty cycle of 98.6 $\%$) before demodulation. See the text for extra details.}
\label{fig:rec}%
\end{figure*}

The restored Stokes I and normalized V are given in Fig. \ref{fig:rec} along with the results of a plain accumulation to show the improvement not only in spatial resolution (compare d and e) but also in SNR (compare g and h) even when the individual intensity measurements have low ($\sim13.9$) SNR (see image b). In addition, we include the outcome of the same restoration algorithm, configured in the same way, to a data set with a modulation frequency of 11.1 Hz (44.4 fps) and duty cycle of 11$\%$, that was derived from the original 100 Hz data as explained in Sect \ref{sec:sicqs}. Note that, in this case, the reduced number of frames available for the restoration ---only 21 instead of 192 per modulation state--- produces a worse restoration (compare e and f) even when the individual intensity measurements have the same exposure time and SNR (compare b and c). 

The restorations presented in  Fig. \ref{fig:sic2} c, d ,e and h; were also done as explained above but with one difference, namely that the modulated intensities were treated as simultaneous objects by the MOMFBD code and thus a single degradation function was fitted per modulation cycle. This minimizes possible polarimetric artifacts introduced by the restoration process.

\section{Conclusion and prospects}
\label{sec:cncl}
We have introduced the instrumental details for the prototype of a novel, high-cadence solar polarimeter. The FSP is designed to obtain high-resolution, high-sensitivity measurements of the full Stokes vector in visible wavelengths from the ground. To achieve this goal the system uses an FLC-based modulator package optimized to have achromatic and balanced polarimetric efficiencies in the 400-700 nm wavelength range (see Fig. \ref{fig:eff}). The intensity detection is done in synchronization with the modulator using a small, custom-made pnCCD camera that can record up to 400 fps and has almost 100$\%$ duty cycle and total noise below 5 $e^-rms$ (see Table \ref{tab:scomp}). 

The high modulation frequency (100 Hz) substantially reduces the levels of SIC artifacts per frame to below the noise in typical, narrow-band measurements (see Fig. \ref{fig:sic} b). The latter is crucial to achieve simultaneously high spatial resolution (by means of post-facto image restoration) and polarimetric sensitivity (see Fig. \ref{fig:sic2}). Moreover, we found no relevant artifact or systematic effect in quiet-Sun measurements, carried out with the modulator package disconnected,  that are above $7\times10^{-5}$ (see Fig. \ref{fig:sic} f). Similar results were obtained for solar scenes with higher contrasts, e.g. active regions.

Due to the low noise and high duty cycle of the pnCCD camera, it can produce photon-noise dominated images for all the most common scientific targets tested so far and thus achieve the desired SNR in the shortest time. The high-cadence and duty cycle of FSP are also beneficial when applying post-facto image restoration. The latter was illustrated by restoring a set of 192 measurements using a MOMFBD, to show that the larger amount of frames available with FSP improves the restoration quality even when the SNR of the individual frames is low (See Fig. \ref{fig:rec}).  

The main drawbacks of the FSP prototype are the small sensor area (to be improved in the second phase of the project) and the relatively large pixel size (see Table \ref{tab:scomp}). The latter may result in large optical setups that are not suitable for instruments with severe space constraints, or that may introduce further image aberrations.

The second phase of the FSP project involves the development of a science-ready instrument. FSPII will use the same modulation package as FSP, described  in Sect. \ref{sec:modchar}, and a new camera currently, under development at the Max Planck Semiconductor Lab. The expected detector specifications are given in Table \ref{tab:scomp}, with the main improvements being the increased sensor area and reduced pixel size. In order to achieve 400 fps with a four times larger sensor, an improved readout ASIC will be used, namely the VERITAS \citep{porro2013}. The expected date for the first-light campaign is 2016. 

\begin{acknowledgements}
  The Fast Solar Polarimeter project is funded by the Max Planck Society (MPG)
  and by the European Commission, grant no. 312495 (SOLARNET). We also would
  like to thank all the technical contributors not listed as co-authors for
  their invaluable input to the project. In particular we would like to thank
  the staff of PNSensor GmbH and of the MPG semiconductor lab in Munich, for
  their excellent work on and their support with the pnCCD camera system, as
  well as the staff of the Kiepenheuer Institut f\"ur Sonnenphysik in
Freiburg for supporting our FSP observing campaigns at the VTT. This work was partly supported by the BK21 plus program through the National Research Foundation (NRF) funded by the Ministry of Education of Korea. The participation of F. A. Iglesias was funded by the International Max Planck Research School for Solar System Science.
\end{acknowledgements}

 \bibliographystyle{aa} 
 \bibliography{iglesias_et_al_2016} 

\end{document}